%% file: main.tex
\documentclass[10pt,conference]{IEEEtran}
\IEEEoverridecommandlockouts

\usepackage{amsmath,amsfonts}
\usepackage{graphicx}
\usepackage{textcomp}
\usepackage{enumitem}
\usepackage{multirow}
\usepackage{colortbl}
\usepackage{array}
\usepackage{tcolorbox}
\usepackage{colortbl}
\usepackage{nicematrix}
\usepackage{tabularx}
\usepackage{balance}
\usepackage{colortbl}
\usepackage{url}
\usepackage[hidelinks]{hyperref}
\usepackage{booktabs}
\usepackage{siunitx}
\usepackage{makecell}
\usepackage{xcolor}

\newcommand{\pval}[2]{%
  \ifnum#1<50 \textbf{\small (#2)} \else \small (#2) \fi
}

\definecolor{lightgray}{gray}{0.9}

\PassOptionsToPackage{normalem}{ulem}
\usepackage{ulem}

\providecolor{added}{rgb}{0,0,1}
\providecolor{deleted}{rgb}{1,0,0}

\usepackage{subcaption}
\usepackage{mwe}

\definecolor{specialblue}{RGB}{0,104,149}

\definecolor{specialgray}{RGB}{242,242,242}

\usepackage{gb4e}
\noautomath

\usepackage{tikz}
\usepackage[framemethod=tikz]{mdframed}
\usetikzlibrary{calc}

\newtcolorbox{mybox}{
  sharp corners,
  colback=specialgray,
  colframe=specialblue,
  boxrule=0pt,
    toprule=0pt,
  bottomrule=0pt,
  leftrule=3pt, 
  rightrule=3pt 
}

\usepackage{cite}
\usepackage{amsmath,amssymb,amsfonts}
\usepackage{algorithmic}
\usepackage{graphicx}
\usepackage{textcomp}
\def\BibTeX{{\rm B\kern-.05em{\sc i\kern-.025em b}\kern-.08em
    T\kern-.1667em\lower.7ex\hbox{E}\kern-.125emX}}
\begin{document}

\title{GUI-ReRank: Enhancing GUI Retrieval with Multi-Modal LLM-based Reranking}

\author{
\IEEEauthorblockN{
Kristian Kolthoff\IEEEauthorrefmark{2}, 
Felix Kretzer\IEEEauthorrefmark{3}, 
Christian Bartelt\IEEEauthorrefmark{2}, 
Alexander Maedche\IEEEauthorrefmark{3},
and Simone Paolo Ponzetto\IEEEauthorrefmark{4}}

\IEEEauthorblockA{
\IEEEauthorrefmark{2}Institute for Software and Systems Engineering,
Clausthal University of Technology, Clausthal, Germany\\
Email: \{kristian.kolthoff, christian.bartelt\} @tu-clausthal.de}

\IEEEauthorblockA{\IEEEauthorrefmark{3}Human-Centered Systems Lab,
Karlsruhe Institute of Technology, Karlsruhe, Germany\\
Email: \{felix.kretzer, alexander.maedche\} @kit.edu}

\IEEEauthorblockA{\IEEEauthorrefmark{4}Data and Web Science Group, 
University of Mannheim, Mannheim, Germany\\
Email: simone@informatik.uni-mannheim.de}}


\maketitle

\begin{abstract}
\input{chapters/00_Abstract}
\end{abstract}


\IEEEpeerreviewmaketitle

\input{chapters/01_Introduction}
\input{chapters/02_Approach}

\input{chapters/03_Evaluation}
\input{chapters/04_Related_Work}
\input{chapters/05_Conclusion}

\ifCLASSOPTIONcaptionsoff
  \newpage
\fi

\bibliographystyle{IEEEtran}
\bibliography{bibtex/bib/ref}

\end{document}

%% file: chapters/00_Abstract.tex
Graphical User Interface (GUI) prototyping is a fundamental component in the development of modern interactive systems, which are now ubiquitous across diverse application domains. GUI prototypes play a critical role in requirements elicitation by enabling stakeholders to visualize, assess, and refine system concepts collaboratively. Moreover, prototypes serve as effective tools for early testing, iterative evaluation, and validation of design ideas with both end users and development teams. Despite these advantages, the process of constructing GUI prototypes remains resource-intensive and time-consuming, frequently demanding substantial effort and expertise. Recent research has sought to alleviate this burden through natural language (NL)-based GUI retrieval approaches, which typically rely on embedding-based retrieval or tailored ranking models for specific GUI repositories. However, these methods often suffer from limited retrieval performance and struggle to generalize across arbitrary GUI datasets. In this work, we present \textit{GUI-ReRank}, a novel framework that integrates rapid embedding-based constrained retrieval models with highly effective multi-modal (M)LLM-based reranking techniques. \textit{GUI-ReRank} further introduces a fully customizable GUI repository annotation and embedding pipeline, enabling users to effortlessly make their own GUI repositories searchable, which allows for rapid discovery of relevant GUIs for inspiration or seamless integration into customized LLM-based retrieval-augmented generation (RAG) workflows. We evaluated our approach on an established NL-based GUI retrieval benchmark, demonstrating that \textit{GUI-ReRank} significantly outperforms state-of-the-art (SOTA) tailored Learning-to-Rank (LTR) models in both retrieval accuracy and generalizability. Additionally, we conducted a comprehensive cost and efficiency analysis of employing MLLMs for reranking, providing valuable insights regarding the trade-offs between retrieval effectiveness and computational resources. Video presentation of \textit{GUI-ReRank} available at: \url{https://youtu.be/_7x9UCh82ug}

%% file: chapters/01_Introduction.tex
\section{Introduction}

A wide range of strategies have been explored to support the elicitation of requirements in user-centric software development \cite{pohl2010requirements}. One particularly impactful technique is GUI prototyping, which helps analysts convey their interpretation of requirements through visual means, while simultaneously offering stakeholders concrete artifacts for review and feedback. Early exposure to functional prototypes not only encourages stakeholder involvement but also promotes richer dialogue and iterative refinement of requirements \cite{ravid2000method, beaudouin2002prototyping}. Nevertheless, leveraging GUI prototyping for requirements gathering often necessitates specialized expertise and entails a significant investment of manual labor. The inherently cyclical and time-consuming nature of prototyping demands repeated interactions between analysts and stakeholders, which can lead to requirements that are quickly outdated or misaligned with evolving stakeholder needs \cite{schneider2007generating}.

Recent advances in LLM-driven GUI generation \cite{chen2021evaluating, brie2023evaluating, kolthoff2024zero, fiebig2025effective} have enabled the automatic creation of user interfaces directly from natural language descriptions, offering remarkable flexibility and the ability to synthesize novel designs. Moreover, these approaches have recently been also integrated into GUI prototyping workflows \cite{kolthoff2024interlinking, kretzer2025closing, kolthoff2025guide}. However, these generative approaches can be computationally expensive and often lack mechanisms for reusing existing high-quality designs, which is particularly valuable for custom prototyping styles or specialized domains not well represented in LLM pretraining. 

\begin{figure*}
\includegraphics[width=\textwidth]{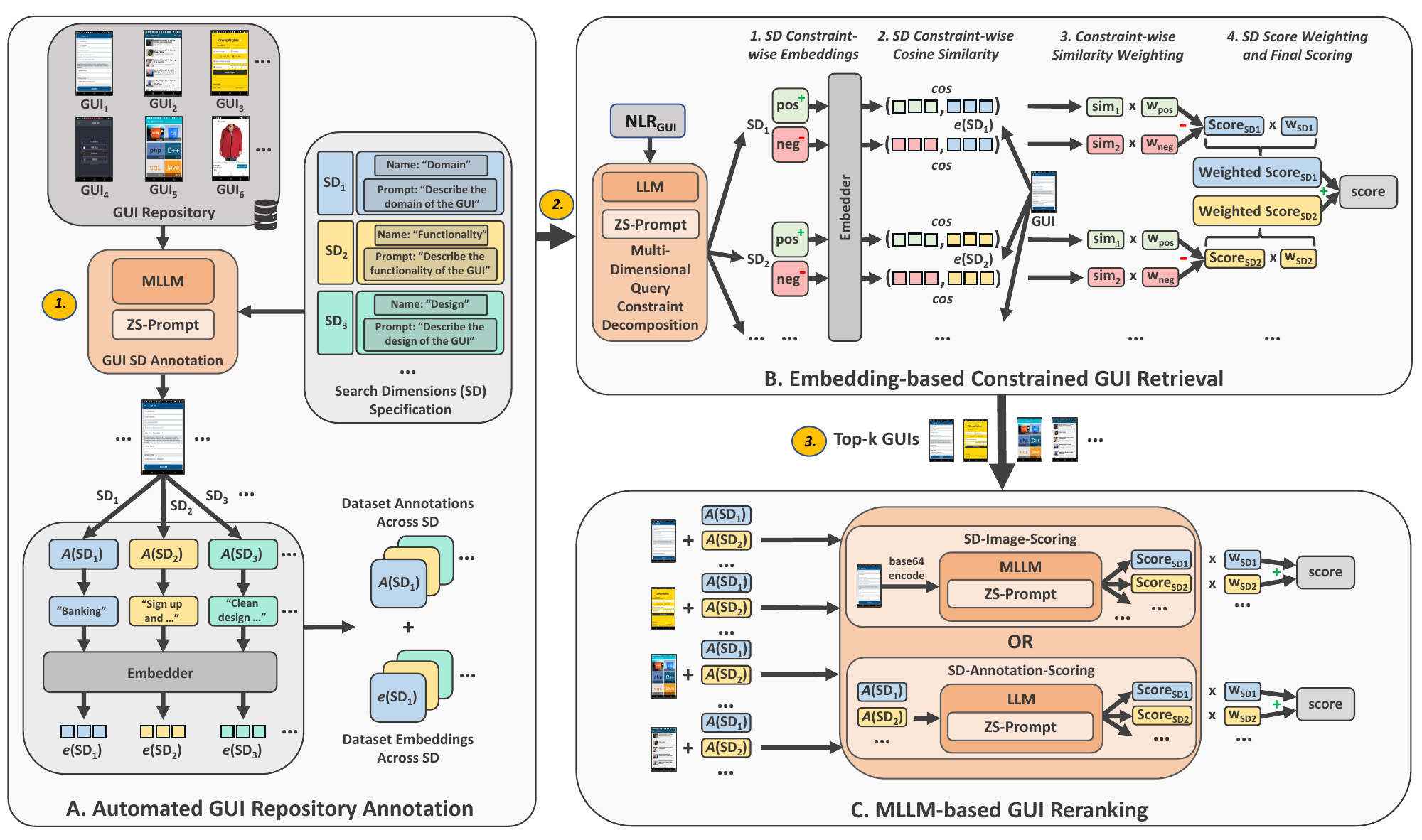}
  \caption{Overview of the \textit{GUI-ReRank} approach with \textit{(A)} the automated GUI repository annotation and embedding pipeline, \textit{(B)} the embedding-based constrained GUI retrieval approach and \textit{(C)} the MLLM-based GUI reranking via images or annotations}
	\label{fig:overview}
\end{figure*}

As a complementary direction, many approaches have been proposed for NL-based GUI retrieval, including systems such as \textit{Guigle} \cite{bernal2019guigle}, \textit{GUI2WiRe} \cite{kolthoff2020gui2wire}, and \textit{RaWi} \cite{kolthoff2023data}, among others \cite{ kolthoff2019automatic, kolthoff2021automated, huang2021creating}. While these GUI retrieval methods facilitate prototyping by rapidly retrieving relevant GUIs from large-scale repositories, they excel at enabling practitioners to efficiently search and reuse vast collections of existing GUIs in a cost-effective manner. Retrieval-based methods can also be seamlessly integrated into LLMs as RAG modules, thereby enhancing generative capabilities for custom domains not covered in pretraining data. Despite these advantages, existing retrieval approaches still exhibit limited performance on benchmark evaluations and are typically tailored to specific GUI datasets, most notably \textit{Rico} \cite{deka2017rico}, by exploiting their unique hierarchical structures. This lack of generalizability and suboptimal retrieval effectiveness highlights the need for more robust, adaptable solutions in NL-based GUI retrieval.

To close this gap, we propose \textit{GUI-ReRank}, a novel framework that significantly advances NL-based GUI retrieval and enables customized search workflows. \textit{GUI-ReRank} employs a two-stage ranking approach, integrating rapid embedding-based retrieval models with powerful multi-modal (M)LLM-based reranking techniques to deliver highly accurate and contextually relevant results. Our approach is designed to support complex constrained natural language requirements (NLR) for search, including negation and multiple customizable dimensions such as \textit{domain}, \textit{functionality}, \textit{design} and \textit{GUI components} already at the embedding-based retrieval stage, enabling more expressive and precise GUI searches than previously possible. Furthermore, \textit{GUI-ReRank} features a fully customizable LLM-based GUI annotation and embedding pipeline, allowing practitioners to seamlessly incorporate custom GUI datasets. \textit{GUI-ReRank} ships with the well-known GUI dataset \textit{Rico} \cite{deka2017rico} pre-annotated and embedded, to provide access to a large-scale and diverse GUI dataset. \textit{GUI-ReRank} empowers developers, designers and analysts (and MLLMs via RAG pipelines) to effectively discover and reuse high-quality GUIs in diverse development contexts. Our source code, datasets and prototype are available at repository \cite{github}.

%% file: chapters/02_Approach.tex
\section{Approach: GUI-ReRank}

\textit{GUI-ReRank} consists of three main components: \textit{(A)} an automated GUI repository annotation module that prepares custom GUI datasets via LLM-based annotation and embedding generation, \textit{(B)} an embedding-based constrained GUI retrieval component, which uses these embeddings combined with LLM-driven query decomposition to extract and utilize positive and negative constraints across multiple search dimensions and computes an initial GUI ranking, and \textit{(C)} an MLLM-based GUI reranking model that performs image- or text-based reranking using zero-shot prompting and computes a weighted score over multiple search dimensions. Fig. \ref{fig:overview} shows an overview of the three components and the overall workflow.

\definecolor{lightgray}{gray}{0.92}

\newcolumntype{G}{>{\columncolor{white}}c}

\begin{table*}[!t]
\footnotesize
\caption{Gold standard results of previous SOTA GUI reranking approaches
         in comparison to MLLM-based GUI reranking}
\centering
\setlength\tabcolsep{5pt}
\renewcommand{\arraystretch}{1.1}

\begin{tabular}{G|l|c|c|cccc|cccc|cccc}
\toprule
\multicolumn{2}{c|}{} & \textbf{AP} & \textbf{MRR} &
        \multicolumn{4}{c|}{\textbf{Precision@k}} &
        \multicolumn{4}{c|}{\textbf{HITS@k}} &
        \multicolumn{4}{c}{\textbf{NDCG@k}} \\
\cmidrule(lr){3-3}\cmidrule(lr){4-4}\cmidrule(lr){5-8}
\cmidrule(lr){9-12}\cmidrule(lr){13-16}
\multicolumn{2}{c|}{} &  AP & MRR & $P@3$ & $P@5$ & $P@7$ & $P@10$
                                    & $H@1$ & $H@3$ & $H@5$ & $H@10$
                                    & $N@3$ & $N@5$ & $N@10$ & $N@15$ \\
\midrule
 & \textbf{BERT-LTR-1}          & 0.486 & 0.618 & 0.377 & 0.350 & 0.307 & 0.269
                                    & 0.460 & 0.710 & 0.860 & 0.980
                                    & 0.530 & 0.560 & 0.634 & 0.697 \\
\rowcolor{lightgray}
\textbf{BL} & \textbf{BERT-LTR-2}          & 0.501 & 0.631 & 0.400 & 0.340 & 0.304 & 0.281
                                    & 0.440 & 0.750 & 0.910 & 0.980
                                    & 0.543 & 0.556 & 0.636 & 0.701 \\
 & \textbf{BERT-LTR-3}          & 0.499 & 0.626 & 0.363 & 0.354 & 0.317 & 0.287
                                    & 0.450 & 0.730 & 0.860 & 1.000
                                    & 0.517 & 0.554 & 0.646 & 0.694 \\
\midrule
    & \textbf{GPT-4.1}               & 0.813 & 0.927 & 0.720 & 0.552 & 0.451 & 0.349
                                    & \textbf{0.870} & \textbf{1.000} & \textbf{1.000} & \textbf{1.000}
                                    & 0.841 & 0.823 & 0.866 & 0.891 \\
\rowcolor{lightgray}
\textbf{Text}     & \textbf{GPT-4.1 Mini}          & 0.797 & 0.923 & 0.673 & 0.546 & 0.446 & 0.344
                                    & 0.870 & 0.980 & 1.000 & 1.000
                                    & 0.817 & 0.810 & 0.844 & 0.880 \\
   & \textbf{GPT-4.1 Nano}          & 0.662 & 0.842 & 0.550 & 0.460 & 0.390 & 0.304
                                    & 0.750 & 0.910 & 0.960 & 1.000
                                    & 0.714 & 0.712 & 0.760 & 0.810 \\
\midrule
    & \textbf{GPT-4.1}               & \textbf{0.840} & 0.928 & \textbf{0.750} & \textbf{0.566} & \textbf{0.463} & \textbf{0.354}
                                    & 0.870 & 0.990 & 1.000 & 1.000
                                    & \textbf{0.877} & \textbf{0.852} & \textbf{0.884} & \textbf{0.908} \\
\rowcolor{lightgray}
\textbf{Image}    & \textbf{GPT-4.1 Mini}          & 0.811 & \textbf{0.930} & 0.717 & 0.540 & 0.441 & 0.351
                                    & 0.880 & 0.980 & 1.000 & 1.000
                                    & 0.850 & 0.820 & 0.868 & 0.895 \\
   & \textbf{GPT-4.1 Nano}          & 0.582 & 0.790 & 0.490 & 0.408 & 0.334 & 0.277
                                    & 0.660 & 0.910 & 0.990 & 1.000
                                    & 0.621 & 0.622 & 0.667 & 0.724 \\
\bottomrule
\end{tabular}
\label{tab:results_rq1}
\end{table*}

\newcolumntype{G}{>{\columncolor{white}}c}

\begin{table}[h]
\footnotesize
\setlength\tabcolsep{4pt}
\renewcommand{\arraystretch}{1.1}
\centering
\begin{tabular}{l l|c|c|c|c|c|c}
\toprule
 &  & \multicolumn{2}{c|}{\textbf{\#Tokens (k=1)}} & \multicolumn{2}{c|}{\textbf{Cost (k)}} & \multicolumn{2}{c}{\textbf{Time (k)}} \\
\cmidrule(lr){3-4} \cmidrule(lr){5-6} \cmidrule(lr){7-8}
 &  & Input & Output & 100 & 500 & 100 & 500 \\
\midrule
  & \textbf{GPT-4.1}       & 179.77 & 6.00  & \$.041 & \$.204  & 2.4s  & 12s \\
\rowcolor{lightgray}
\textbf{Text}  & \textbf{GPT-4.1 Mini}  & 179.77 & 6.00  & \$.008 & \$.041  & 2.6s   & 13s \\
  & \textbf{GPT-4.1 Nano}  & 179.77 & 7.11  & \$.002 & \$.010  & 2s   & 10s \\
\midrule
 & \textbf{GPT-4.1}       & 1089.02 & 6.00  & \$.223 & \$1.113  & 12.4s  & 62s \\
\rowcolor{lightgray}
\textbf{Image} & \textbf{GPT-4.1 Mini}  & 2154.96 & 6.00  & \$.087 & \$.436  & 9s  & 45s \\
 & \textbf{GPT-4.1 Nano}  & 3242.41 & 5.95  & \$.033 & \$.163  & 8s  & 40s \\
\bottomrule
\end{tabular}
\caption{Token usage, cost and runtime for \textit{GPT-4.1} models (denotes means computed over the gold standard runs, \textit{k}=number of retrieved GUIs to rerank, \#\textit{Celery} workers = 10}
\label{tab:token_cost_runtime}
\end{table}

\subsection{Automated GUI Repository Annotation}

Given a GUI repository as images, \textit{GUI-ReRank} initiates the process by employing a zero-shot prompted MLLM to generate annotations for each GUI based on a customizable set of search dimensions (SD). The framework provides a default set of SD, including \textit{domain}, \textit{functionality}, \textit{design}, \textit{GUI components}, and \textit{displayed text}, which are broadly applicable for many GUI retrieval scenarios. However, users can readily extend or modify these dimensions to address specific requirements, e.g., \textit{platform} or \textit{accessibility}, by specifying a name and an NL description for each SD. Subsequently, each annotation is processed by an embedding model to produce SD-specific embeddings, which, together with the annotations, form the foundation for efficient and flexible search across the dataset. 

\subsection{Embedding-based Constrained GUI Retrieval}

To support complex semantic searches with both positive and negative constraints across multiple search dimensions, \textit{GUI-ReRank} first decomposes the natural language requirements (NLR) using a zero-shot prompted LLM, extracting positive and negative constraints for each SD (e.g., $SD_{Design}$: $pos^{+}("modern")$, $neg^{-}("dark")$). For each dimension, we compute the cosine similarity between the user-specified constraints and the corresponding GUI annotations, quantifying the degree of match for both positive and negative aspects. These similarity scores are then weighted and the negative similarity is subtracted from the positive, thus a higher negative match proportionally reduces the overall positive score for each dimension. The final retrieval score for each GUI is computed as the normalized sum of the weighted similarities across all dimensions, allowing for user-defined emphasis on specific SD. This approach enables nuanced, constraint-based retrieval such as explicitly excluding certain features directly during initial and cost-effective embedding-based retrieval, overcoming the limitations of standard embedding models and thus supporting flexible, user-driven NL-based GUI retrieval.

\subsection{Multi-Modal (M)LLM-based GUI Reranking}

Building on the top-\textit{k} GUIs retrieved in the previous stage, the multi-modal LLM-based GUI reranking component refines the results by evaluating either the GUI images or their textual SD annotations using zero-shot prompted MLLMs. For each search dimension, the MLLM assigns a score between 0 (\textit{no match}) and 100 (\textit{perfect match}), reflecting the relevance of each GUI to the NLR of the user. \textit{GUI-ReRank} offers support for both text-based and image-based reranking: text-based reranking offers greater speed and cost-efficiency by leveraging abstracted SD annotations, while image-based reranking provides the most detailed assessment at the expense of higher computational cost and latency. Similar to the previous stage, the dimension-specific scores are weighted according to user preferences, summed and normalized to produce the reranking.

\subsection{Prototype Implementation}

The \textit{GUI-ReRank} prototype is implemented as a \textit{Django} web application with a \textit{HTML/CSS} and \textit{JavaScript} frontend combined with \textit{MySQL} for data storage. Background processing tasks, such as large-scale GUI annotations, embedding generation and parallelized batch-wise reranking of top-\textit{k} GUIs are managed using \textit{Celery} distributed across multiple worker nodes, with \textit{Redis} serving as the message broker to ensure efficient and scalable task execution. Our current prototype supports a range of widely employed LLMs, including \textit{GPT} from \textit{OpenAI}, \textit{Google Gemini} and \textit{Anthropic Claude Sonnet}, allowing users to select the most suitable model for their needs.

%% file: chapters/03_Evaluation.tex
\section{Experimental Evaluation}

In the following, we present a comprehensive evaluation of the \textit{GUI-ReRank} approach, comparing its effectiveness and efficiency against SOTA NL-based GUI ranking approaches.

\subsection{Experimental Setup}

We evaluated our GUI reranking models with MLLMs possessing different cost/performance trade-offs (\textit{GPT-4.1}, \textit{GPT-4.1 Mini} and \textit{GPT-4.1 Nano} (accessed in \textit{July 2025}, \textit{temperature=0.05}) using an established gold standard for NL-based GUI ranking \cite{kolthoff2023data}. This benchmark comprises 100 NL queries, each paired with 20 GUIs from \textit{Rico} \cite{deka2017rico} and enables direct comparison with state-of-the-art \textit{BERT-LTR} models \cite{kolthoff2023data}. We report \textit{Average Precision (AP)}, \textit{Mean Reciprocal Rank (MRR)}, \textit{Precision@}\textit{k}, \textit{HITS@}\textit{k} and \textit{NDCG@}\textit{k}, similar to previous work. To assess efficiency, we report \textit{input/output token consumption} (per reranked GUI or \textit{k=1}), \textit{runtime} and API \textit{costs} (both for \textit{k=100} and \textit{k=500}) for each reranking configuration.

\subsection{Evaluation Results}

Table~\ref{tab:results_rq1} presents the ranking performance of our MLLM-based GUI reranking models compared to previous SOTA \textit{BERT-LTR} approaches. Both text- and image-based reranking models substantially outperform \textit{BERT-LTR} baselines across all metrics. Comparing image-based and text-based models, we observe a higher effectiveness for image-based models. Moreover, the observed effectiveness of the models generally aligns with their performance profiles, with \textit{GPT-4.1} achieving the highest scores, closely followed by the \textit{Mini} model, which, however, is substantially more cost-effective. Table~\ref{tab:token_cost_runtime} summarizes the efficiency analysis. Text-based reranking is significantly more cost- and time-efficient in contrast to the image-based GUI reranking, with token usage and inference cost for \textit{k=100} and \textit{k=500} GUIs significantly reduced. Furthermore, when comparing text-based \textit{GPT-4.1} against the next best model \textit{GPT-4.1 Mini}, only 1.9\% of \textit{Average Precision} performance is lost, while reducing the cost by 79.9\%. These results provide practical guidance for selecting the appropriate reranking mode and model based on resource constraints and application requirements.  Overall, \textit{GUI-ReRank} delivers SOTA GUI reranking performance with flexible efficiency trade-offs.

%% file: chapters/04_Related_Work.tex
\section{Related Work}

A variety of methods have been proposed for NL-based GUI retrieval to facilitate rapid prototyping and reuse of interface designs. \textit{Guigle} \cite{bernal2019guigle} introduced one of the first search engines for GUI screenshots, leveraging traditional information retrieval techniques such as \textit{TF-IDF} and \textit{BM25} over multiple facets of the GUI hierarchy. Other approaches have utilized pooled \textit{BERT} embeddings and multi-modal embedding spaces to enhance text-based GUI retrieval \cite{huang2021creating}. Advanced retrieval methods and \textit{BERT-based Learning-to-Rank} (\textit{LTR}) models have also been adopted in approaches like \textit{GUI2WiRe} \cite{kolthoff2020gui2wire, kolthoff2021automated} and \textit{RaWi} \cite{kolthoff2023data}, which exploit the hierarchical structure of datasets such as \textit{Rico} \cite{deka2017rico} to improve retrieval accuracy. Beyond general GUI retrieval, \textit{d.tour} \cite{ritchie2011d} focuses on stylistic similarity and keywords to find websites with comparable design aesthetics, while \textit{Gallery D.C.} \cite{chen2019gallery} extracts GUI components from real-world applications and provides a multi-modal search engine supporting queries across dimensions such as \textit{size}, \textit{color} and \textit{text}. Although these retrieval methods enable efficient search and reuse of large-scale GUI repositories, they often show limited performance on benchmarks and are typically tailored to specific datasets such as \textit{Rico} \cite{deka2017rico}, restricting their generalizability. In contrast, \textit{GUI-ReRank} integrates rapid embedding-based constrained retrieval methods with advanced multi-modal LLM-based reranking and a flexible GUI dataset annotation and embedding pipeline. This combination achieves substantially higher retrieval accuracy and robustness as demonstrated in our evaluation, while also supporting customizable search dimensions and seamless adaptation to proprietary or domain-specific GUI repositories.

%% file: chapters/05_Conclusion.tex
\section{Conclusion}

In this work, we introduced \textit{GUI-ReRank}, a novel framework that combines embedding-based constrained retrieval over multiple customizable search dimensions with MLLM-based reranking to achieve state-of-the-art performance in NL-based GUI ranking. Our experiments show substantial gains over previous GUI retrieval methods and offer valuable insights into efficiency trade-offs between the different models.